\documentclass[pra,superscriptaddress,nofootinbib]{revtex4}
\usepackage{graphicx}
\usepackage{amsmath}
\usepackage{amsthm}
\usepackage{amssymb}
\usepackage{float}
\usepackage{algorithm}

\floatname{algorithm}{Protocol}
\usepackage{bm}
\usepackage{bbm}

\usepackage{color}

\def\EQ#1{\begin{eqnarray}#1\end{eqnarray}}
\newcommand{\djj}{d\kern-0.4em\char"16\kern-0.1em}

\newtheorem{corollary}{Corollary}

\newtheorem{defn}{Definition}
\newtheorem{prop}{Proposition}\def\PRO{\begin{prop}}\def\ORP{\end{prop}}
\newtheorem{coro}{Corollary}\def\COR{\begin{coro}}\def\ROC{\end{coro}}
\newtheorem{theo}{Theorem}\def\TH{\begin{theo}}\def\HT{\end{theo}}
\def\TH{\begin{theo}}\def\HT{\end{theo}}
\newtheorem{defi}[prop]{Definition}\def\DE{\begin{defi}}\def\ED{\end{defi}}
\newtheorem{thm}{Theorem}
\newtheorem{obs}[thm]{Observation}
\newtheorem{lemme}[prop]{Lemma}\def\LE{\begin{lemme}}\def\EL{\end{lemme}}
\def\ket#1{\left| #1 \right\rangle}

\newcommand{\beq}{\begin{equation}}
\newcommand{\eeq}{\end{equation}}

\parskip=\medskipamount

\begin{document}
\title{Multiparty Quantum Signature Schemes}
\author{Juan Miguel Arrazola}
\affiliation{Institute for Quantum Computing, University of Waterloo, 200 University Avenue West, Waterloo, Ontario, N2L 3G1, Canada}
\author{Petros Wallden}
\affiliation{School of Informatics, University of Edinburgh,
10 Crichton Street, Edinburgh EH8 9AB, UK}
\author{Erika Andersson}
\affiliation{SUPA, Institute of Photonics and Quantum Sciences,
Heriot-Watt University, Edinburgh, EH14 4AS, UK}

\begin{abstract}
Digital signatures are widely used in electronic communications to secure important tasks such as financial transactions, software updates, and legal contracts. The signature schemes that are in use today are based on public-key cryptography and derive their security from computational assumptions. However, it is possible to construct unconditionally secure signature protocols. In particular, using quantum communication, it is possible to construct signature schemes with 
security based on fundamental principles of quantum mechanics. Several quantum signature protocols have been proposed, but none of them has been explicitly generalised to more than three participants, and their security goals have not been formally defined. Here, we first extend the security definitions of Swanson and Stinson \cite{swanson2011unconditionally} so that they can apply also to the quantum case, and introduce a formal definition of transferability based on different verification levels. 
We then prove several properties that multiparty signature protocols with information-theoretic security -- quantum or classical -- must satisfy in order to achieve their security goals. We also express two existing quantum signature protocols with three parties in the security framework we have introduced. Finally, we generalize a quantum signature protocol given in \cite{dunjko2014QDSQKD} to the multiparty case, proving its security against forging, repudiation and non-transferability. Notably, this protocol can be implemented using any point-to-point quantum key distribution network and therefore is ready to be experimentally demonstrated.
\end{abstract}

\maketitle

\section{Introduction}

Digital signatures are important cryptographic building blocks which are widely used to provide security in electronic communications. They achieve three main cryptographic goals: authentication, non-repudiation, and transferability. These properties make them suitable for securing important tasks such as financial transactions, software updates, and legal contracts. The digital signatures schemes that are in use today, which are based on public-key cryptography, derive their security from unproven computational assumptions, and most of them -- notably those based on the RSA algorithm or on elliptic curves -- can be broken 
by quantum computers~\cite{childs2010quantum}.

Consequently, from both a practical and fundamental perspective, there has been an interest in studying 
signature protocols that do not rely on computational assumptions, but instead offer information-theoretic security. These schemes were first introduced by Chaum and Roijakkers \cite{chaum1991unconditionally} and are known as \textit{unconditionally secure signature} (USS) schemes. Besides the proposal of Chaum and Roijakkers, several other USS protocols have been suggested \cite{brickell1988authentication,hanaoka2000unconditionally,hanaoka2002efficient,
johansson1994construction,johansson1999further,safavi2004general,simmons1988message,simmons1990cartesian,dunjko2014QDSQKD}, most of them based on removing standard trust assumptions from message authentication codes (MACs). However,  most known classical USS protocols proposed so far rely on the assumption of either a trusted arbiter or authenticated broadcast channels, and crucially, all of them require the use of secure channels, which are impossible to realize, practically, with information-theoretic security using only classical communication~\cite{shannon1949communication,maurer1993secret}. 

Once quantum communication is allowed, it becomes possible to construct l 
signature schemes whose information-theoretic security is based on fundamental principles of quantum mechanics. These are known as quantum signature (QS) schemes. The first QS protocol was proposed by Gottesman and Chuang \cite{QDS}, who introduced the main ideas for bringing digital signatures into the quantum world. Although influential from a fundamental point of view, their scheme requires the preparation of complex quantum states, performing quantum computation on these states and storing them in quantum memory, making the protocol highly impractical. This is also an issue for other protocols that appeared shortly after \cite{muller2002quantum,lu2005quantum}.

Recently, new QS protocols that do not require a quantum memory and which can be realized with standard quantum-optical techniques have been proposed \cite{dunjko2014QDSQKD,dunjko2014quantum,arrazola2014quantum}. Some of these protocols have also been demonstrated experimentally \cite{clarke2012experimental,CollinsQDS}, thus establishing their viability as a practical technology. A short review of these developments can be found in Ref. \cite{AA15}. Nevertheless, these schemes have not been generalized to more than three participants, and their security goals have not been formally defined. Overall, a security framework for quantum signature schemes that includes rigorous definitions of security suitable for multiparty protocols has not yet been proposed. In the absence of such a framework, it is not clear how to design secure multiparty protocols nor what the concrete advantages of quantum 
signatures are compared to their classical counterparts.

In this work, we provide a security framework suitable for USS protocols involving an arbitrary number of participants. We follow the definitions by
Swanson and Stinson \cite{swanson2011unconditionally}, generalizing them so that they can apply also to the quantum case, and introduce a formal definition of transferability based on different verification levels. We also present a characterization of the general structure of USS protocols and introduce rigorous definitions of security. Additionally, we prove several properties that these protocols must satisfy in order to achieve their security goals. We then express two existing protocols for quantum signatures with three parties within the framework we developed. Finally, we make use of our results to generalize a quantum protocol of Wallden et. al \cite{dunjko2014QDSQKD} to the multiparty case and prove its security against forging, repudiation and non-transferability. Notably, this protocol can be implemented using any point-to-point quantum key distribution network and therefore is ready to be experimentally demonstrated.

\section{Definitions for USS Protocols}\label{Defns}

A QS protocol is carried out by a set of participants and is divided into two stages: the \textit{distribution} stage and the \textit{messaging} stage. The distribution stage is a quantum communication stage, where the parties exchange quantum and classical signals according to the rules of the protocol. Although in principle they could store the received quantum states in a quantum memory, we focus on more practical protocols in which the participants perform measurements on the states and store the outcomes in a classical memory. The participants may also process their data and communicate classically with each other. Overall, each participant is left with a set of rules for signing messages and for verifying signatures. These rules generally depend on their measurement outcomes and the classical communication. At the end of the distribution stage, the parties decide whether to continue to the messaging stage or to abort the protocol. In the messaging stage, one of the participants (the signer) signs a message by attaching a classical string (the signature) to the message. When a participant receives a signed message, they verify its validity according to the rules of the protocol.

A USS protocol must achieve authenticity, non-repudiation, and transferability as its main security goals. Informally, these goals can be defined as follows:

\begin{enumerate}
\item \underline{Authentication:} Except with negligible probability, an adversary cannot create a message and signature pair that is accepted by another participant, i.e. a signature cannot be forged. 
\item \underline{Non-repudiation:} Except with negligible probability, a signer cannot later successfully deny having signed a message that has been accepted by an honest recipient.
\item \underline{Transferability:} A recipient that accepts a signed message can be confident that, except with negligible probability, the signature will also be accepted by other participants.  
\end{enumerate}

In order to satisfy non-repudiation and transferability, each recipient must have a method of determining whether other participants will also agree on the validity of a signature. This is straightforward in classical public-key schemes, since every recipient applies the same rule to verify a signature. However, as we discuss later in this paper, in an information-theoretic scenario, every recipient must have a different rule to verify a signed message -- or, at least, two participants must have the same verification algorithm with low probability\footnote{Following Swanson and Stinson~\cite{swanson2011unconditionally}, with ``verification algorithm'', we understand a full specification of the rules an individual participant is using to verify a message. For example, different recipients could use a more generic ``verification function'', which is the same for all participants, but with random inputs which differ for different recipients. What we mean by the verification algorithm of an individual participant would, in this case, be the generic verification function together with that participant's specific random inputs. This way of defining the recipients' verification functions makes sense considering that a recipient might in this example know neither what what the underlying generic function is, nor what the random inputs are, only what the resulting combination of the generic verification function and random inputs is. This definition of verification function also makes sense for quantum signature protocols.}. Thus, a security model for USS schemes must deal carefully with the notion of non-repudiation and the transferability of signatures.

We now generalize the work of Swanson and Stinson \cite{swanson2011unconditionally} in the context of USS schemes to construct formal definitions that are also suitable for quantum signature schemes and allow for different levels of verification. This will permit us to formalize the structure of general USS protocols, provide rigorous security definitions, and illustrate properties they must possess in order to be secure.

{\defn 
A USS protocol $\mathcal{Q}$ is an ordered set $\{\mathcal{P},X,\Sigma,L,\mathrm{Gen}, \mathrm{Sign}, \mathrm{Ver}\}$ where:

\begin{itemize}

\item[-] The set $\mathcal{P}=\{P_0,P_1,\ldots,P_{N-1}\}$, is the set of $N$ different participants involved in the protocol. We fix $P_0$ to be the signer, and $P_i$ are the possible recipients, with $i\in\{1,\cdots,N-1\}$. $X$ is the set of possible messages and $\Sigma$ is the set of possible signatures.

\item[-] $\mathrm{Gen}$ is the generation algorithm that gives rise to the functions $\mathrm{Sign}$ and $\mathrm{Ver}$ that are used to generate a signature and verify its validity. More precisely, the generation algorithm specifies the instructions for the quantum and classical communication that takes place in the distribution stage of the protocol. Additionally, the generation algorithm instructs how to construct the functions $\mathrm{Sign}$ and $\mathrm{Ver}$ based on the data obtained during the distribution stage. The generation algorithm includes the option of outputting an instruction to abort the protocol.

\item[-] The signature function $\mathrm{Sign}$ is a deterministic function $X\rightarrow\Sigma$ that takes a message $x$ and outputs a signature $\sigma\in\Sigma$.

\item[-] $L=\{-1,0,1,\ldots,l_{\max}\}$ is the set of possible verification levels of a signed message. A verification level $l$ corresponds to the minimum number of times that a signed message can be transferred sequentially to other recipients. For a given protocol, the maximum number of sequential transfers that can be guaranteed is denoted by $l_{\max}\leq N-1$.

\item[-] The verification function $\mathrm{Ver}$ is a deterministic function $X\times \Sigma\times\mathcal{P}\times L\rightarrow\{\mathrm{True,False}\}$ that takes a message $x$, a signature $\sigma$, a participant $P_i$ and a level $l$, and gives a truth value depending on whether participant $P_i$ accepts the signature as valid at the verification level $l$. We denote a verification function with a fixed participant $P_i$ and level $l$ as $\mathrm{Ver}_{i,l}(x,\sigma):=\mathrm{Ver}(x,\sigma,i,l)$.

\end{itemize}}

\begin{figure}
\includegraphics[width=0.7\columnwidth]{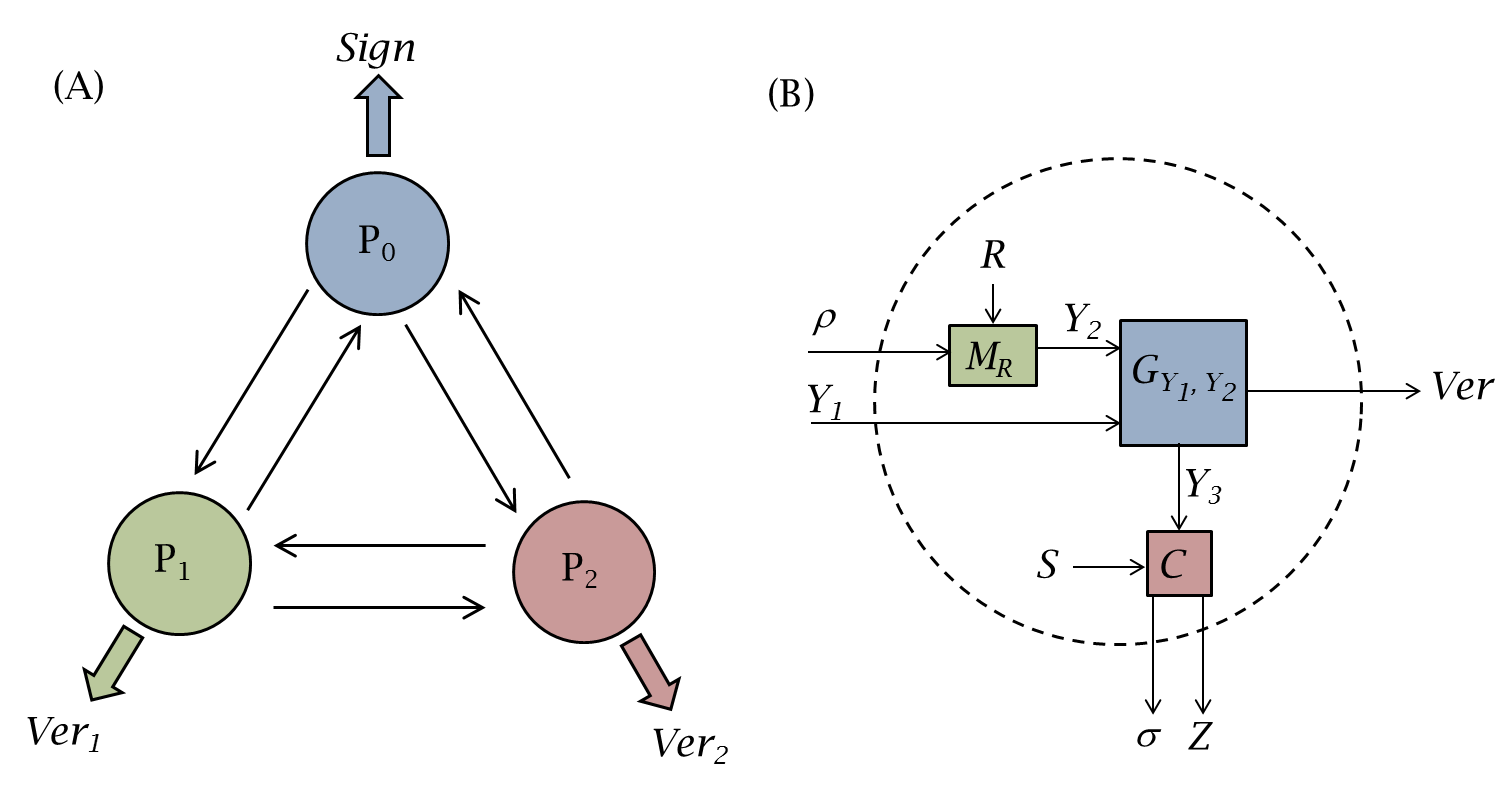}
\caption{(A) Schematic portrayal of a possible generation algorithm in the distribution stage of a QS protocol with three participants. The three parties exchange messages over classical and quantum channels. At the end of their communication, the signer has a specification of the signing algorithm, and the recipients have a specification of their respective verification functions. (B) An example of a generation algorithm for one of the recipients. From their perspective, they receive a quantum state $\rho$ and a classical message $Y_1$ from the other participants. A measurement $M_R$ that depends on a random variable $R$ is carried out on the quantum state, and the outcome $Y_2$, together with the classical data $Y_1$, is fed to an algorithm $G_{Y_1,Y_2}$. This program outputs data $Y_3$ that, together with another possibly random variable $S$, is fed to a second algorithm $C$ that determines the quantum and classical messages sent to the other participants. After several iterations of these steps, the program $G_{Y_1,Y_2}$ outputs the verification function.}\label{Gen}
\end{figure}

In general, the generation algorithm must 
involve randomness in the construction of the signing and verification functions. The randomness may be generated locally by each participant or it can be generated and distributed by a trusted third party. It can arise from the intrinsic randomness of quantum measurements, or by other means. Therefore, even though the signing and verification functions are deterministic functions, they are randomly generated. An illustration of the distribution stage for a generic USS protocol can be seen in Fig. \ref{Gen}.

The verification levels are a method of determining whether a signature can be transferred sequentially among participants. As an illustration, consider a protocol involving a signer Alice, a recipient Bob, and a bank. Other participants may be involved as well. Bob receives a payment from Alice which is signed using a USS protocol, and Bob wants to transfer this signed message to the bank. For Bob, it does not suffice to verify that the signature comes from Alice and that she cannot repudiate it -- he also needs a guarantee that when he transfers the signed message to the bank, they will be able to validate it. Now suppose that the bank also requires the ability to transfer the message to another participant, otherwise they don't accept the message. Then Bob needs a guarantee that it can be transferred \emph{twice} in sequence, from himself to the bank and from the bank to another participant. In general, Bob may require that a signed message be transferred many times in sequence. This guarantee is provided by the verification levels: With high probability, a signature that is verified at level $l$ can be transferred $l$ times in sequence. A signature that is verified at level $l=0$ is certified to have come from the signer, but does not have a guarantee that it can be transferred to other participants. The role of the verification level $l=-1$ is to prevent repudiation, as will be explained in section \ref{Defns}.

We now introduce additional useful definitions, which are inspired by Ref. \cite{swanson2011unconditionally} and generalized to allow different levels of verification. As a starting point, it is important that a USS protocol works properly when all parties are honest.

{\defn A USS protocol $\mathcal{Q}$ is \emph{correct} if $\mathrm{Ver}_{(i,l)}(x,\mathrm{Sign}(x))=\mathrm{True}$  for all $x,i,l$.}
 
Since USS protocols have different verification functions for different participants 
as well as different levels of verification, it is important to carefully specify what it means for a particular signature to be valid.

{\defn A signature $\sigma$ on a message $x$ is \emph{authentic} if $\sigma=\mathrm{Sign}(x)$. }

{\defn A signature $\sigma$ on a message $x$ is \emph{valid} if $\mathrm{Ver}_{(i,0)}(x,\sigma)=\mathrm{True}$ for all $i\in\{1,\cdots,N-1\}$.}

Thus, a valid signature is simply one for which all participants can verify that it originates from the intended signer.  Crucially, a valid signature does not need to be authentic, a possibility not originally considered in Ref. \cite{swanson2011unconditionally}.

{\defn A signature $\sigma$ on a message $x$ is \emph{$i$-acceptable} if $\mathrm{Ver}_{(i,0)}(x,\sigma)=\mathrm{True}$.}

Note that, as opposed to a valid signature, an $i$-acceptable signature may not pass the verification functions of participants other than $P_i$. Therefore, an $i$-acceptable signature may not be a valid signature.

{\defn A signature $\sigma$ on a message $x$ is \emph{$i$-fraudulent}, if $\sigma$ is $i$-acceptable but not valid.}

As discussed before, the participants may additionally be interested in the transferability of the signature. This motivates the following definitions.

{\defn A signature $\sigma$ on a message $x$ is \emph{$l$-transferable} if $\mathrm{Ver}_{(i,l)}(x,\sigma)=\mathrm{True}$ for all $i\in\{1,\cdots,N-1\}$ and there exists $j$ such that $\mathrm{Ver}_{(j,l+1)}(x,\sigma)=\mathrm{False}$. For $l=l_{\max}$, the function $\mathrm{Ver}_{(j,l_{\max}+1)}(x,\sigma)$ is not defined and we assume by convention that it is always $\mathrm{False}$.}

The above definition means that a signature is $l$-transferable if $l$ is the largest level for which this signature will pass the verification test of all participants.

{\defn A signature $\sigma$ on a message $x$ is \emph{$(i,l)$-transferable} if $\mathrm{Ver}_{(i,l)}(x,\sigma)=\mathrm{True}$ and $\mathrm{Ver}_{(i,l+1)}(x,\sigma)=\mathrm{False}$.}

Thus, an $(i,l)$-transferable signature will pass the verification test of participant $i$ at level $l$, but not at any other higher level. As opposed to an $l$-transferable signature, it may not pass the verification functions of other participants.

\subsection{Dispute Resolution}

In traditional digital signature schemes based on public-key cryptography, there is a public verification function to test the validity of a signature. If a person denies having signed a message, the recipient who initially verified the signature can show it to other honest parties -- a judge for example -- who will use the same public verification function to certify its validity and therefore reject the signer's claims. 

However, as we show in section \ref{Properties}, in a USS scheme different participants have different verification functions, 
which makes it possible in principle for two or more participants to disagree on the validity of a signature. The mechanism to prevent repudiation must take this into account.  
Suppose that Alice signs a contract and sends it to Bob, who uses his verification function to verify the signature. The signature passes his verification test at level $l=0$ and he is convinced that the message comes from Alice. Later, Alice attempts to repudiate by denying that she signed the contract. Bob knows that the other participants have different verification functions than his own, so what can be done
to prevent Alice from repudiating? Non-repudiation is ensured by incorporating a dispute resolution method: a mechanism to handle the event of a disagreement on the validity of a signature.
 It is expected that dispute resolution will be invoked relatively rarely. It is akin to an appeal procedure which is very expensive for the participant who loses. Any participant (honest or dishonest), who thinks he might lose the dispute resolution, will avoid any action that could lead to someone invoking it. Therefore, while the dispute resolution may seem complicated and resource-expensive in terms of communication, is not something that affects the effectiveness of the protocol, since any rational participant, whether adversarial or honest, will always take actions that guarantee that any other rational participant would not invoke dispute resolution. 
 Based on Ref. \cite{swanson2011unconditionally}, we formally define such a dispute resolution method as follows.

{\defn A dispute resolution method $\mathrm{DR}$ for a USS scheme $\mathcal{Q}$ is a procedure invoked whenever there is a disagreement on whether a signature $\sigma$ on a message $x$ is a valid signature originating from the signer $P_0$. The participant invoking the dispute resolution can be anyone, including the signer $P_0$. The procedure consists of an algorithm $\mathrm{DR}$ that takes as input a message-signature pair $(x,\sigma)$ and outputs a value $\{$\emph{Valid, Invalid}$\}$ together with the rules:

\begin{enumerate}
\item If $\mathrm{DR}(x,\sigma)$ outputs \emph{Valid}, then all users must accept $(x,\sigma)$ as a valid signature for $x$.
\item If $\mathrm{DR}(x,\sigma)$ outputs \emph{Invalid}, then all users must reject $(x,\sigma)$ as a valid signature for $x$.
\end{enumerate}}

Defining a particular dispute resolution method constitutes a crucial part of specifying a USS protocol. Whether a protocol is secure against repudiation will generally depend on the choice of dispute resolution. But what are the concrete possibilities that we can choose from? A simple strategy is to designate a trusted participant to be in charge of deciding the validity of a signature whenever the dispute resolution method is invoked. This participant, who may have access to more information about the protocol than others, serves as an arbiter who has the final word whenever there is a dispute. An obvious drawback of this choice is the necessity of trust: If the arbiter behaves dishonestly, perhaps due to being blackmailed to do so, the entire security of the protocol is compromised. In this paper, we focus on a \textit{majority vote} dispute resolution method.

{\defn When the validity of a message-signature pair $(x,\sigma)$ is invoked, a majority vote dispute resolution method $\mathrm{MV}(x,\sigma)$ is defined by the following rule:
\begin{enumerate}
\item $\mathrm{MV}(x,\sigma)=$ \emph{Valid} if $\mathrm{Ver}_{(i,-1)}(x,\sigma)=\mathrm{True}$ for more than half of the users.
\item $\mathrm{MV}(x,\sigma)=$ \emph{Invalid} otherwise,
\end{enumerate}
where $\mathrm{Ver}_{(i,-1)}$ is the verification function at level $l=-1$.}

The need for a verification level $l=-1$ can be understood as a mechanism to prevent repudiation by Alice, and it is only relevant when $\mathrm{DR}$ is invoked. Intuitively, $\mathrm{Ver}_{(i,-1)}$ should be chosen such that is infeasible to produce a signature that passes the verification function of one participant at level $l=0$, but does not pass the verification function of the majority of participants at level $l=-1$. This will be formalized in section \ref{Properties}.

The majority vote dispute resolution method was implicitly used in the protocols of \cite{dunjko2014QDSQKD,dunjko2014quantum} when discussing security against repudiation. The obvious advantage of the majority vote method is that we do not need to trust any fixed participant, but instead assume only that at least \emph{most} of them are not dishonest. However, we emphasize that the security definitions of the following section do not depend on a particular choice of $\mathrm{DR}$.

Note that a dispute resolution method can be used by any participant to convince others of the validity of a signature, even when the signature is only verified at level $l=0$. If the protocol is secure against repudiation -- as will be formally defined in the next section -- then no person other than the signer will be able to create a signature that is deemed valid by the dispute resolution method. Therefore, if $\mathrm{DR}$ is invoked and outputs ``Valid", everyone is already convinced that the signature must have come from the signer. This means that the verification levels serve the specific purpose of providing the participants with an assurance that other people will sequentially verify a transferred signature \emph{without the need to invoke dispute resolution.} This is desirable because carrying out dispute resolution may be expensive and should only be invoked under special circumstances.

Finally, we also consider the case in which a participant is dishonest about the level at which they verify a signature. For instance, suppose that Bob wants to transfer a message regarding a payment by Alice, signed by Alice, to a store. The store only accepts signatures that they can transfer to a bank, so Bob needs an assurance that Alice's signature can be transferred twice in sequence. Bob verifies the signature at level $l=2$ and sends it to the store. The store, however, is dishonest, and lies to Bob by claiming that they verified the signature only at level $l=0$, even though Bob knows that they should have verified it at least at level $l=1$. If the protocol is secure against repudiation, Bob can invoke dispute resolution to make everyone, including the bank, agree on the validity of the signature. But in order to resolve disputes regarding the verification level of a signature, we need an additional dispute resolution method. 

{\defn A transferability dispute resolution method at level $l$, $\mathrm{TDR}$, for a QS scheme $\mathcal{Q}$, consists of an algorithm $\mathrm{DR_l}$ that takes as input a message-signature pair $(x,\sigma)$ and verification level $l$ and outputs $\{$\emph{$l-$transferable, not $l-$transferable}$\}$ together with the rules:

\begin{enumerate}
\item If $\mathrm{DR_l}(x,\sigma,l)$ outputs \emph{$l-$transferable}, then all users must accept $(x,\sigma)$ as an $l$-transferable signature for $x$.
\item If $\mathrm{DR_l}(x,\sigma,l)$ outputs \emph{not $l$-transferable}, then all users must reject $(x,\sigma)$ as an $l$-transferable signature for $x$.
\end{enumerate}}

For this form of dispute resolution method, we can also use a majority vote method defined as before. 

{\defn A majority vote transferability dispute resolution method at level $l$, $\mathrm{MV}(x,\sigma,l)$, is defined by the following rule:
\begin{enumerate}
\item $\mathrm{MV}(x,\sigma,l)=$ \emph{$l-$transferable} if $\mathrm{Ver}_{(i,l-1)}(x,\sigma)=\mathrm{True}$ for more than half of the users.
\item $\mathrm{MV}(x,\sigma,l)=$ \emph{not $l-$transferable} otherwise.
\end{enumerate}}

If the protocol offers transferability, as will be formally defined in the next section, any participant who verifies a signature at level $l$ has a guarantee that, with high probability, any other participant will verify the signature at level at least $l-1$. Therefore, if the majority of participants are honest, a majority vote will indeed deem the signature that was verified at level $l$ by an honest participant as an $(l-1)$-transferable signature. This form of dispute resolution can serve as a deterrent for dishonest behaviour. In our previous example, the store is discouraged from lying to Bob as they know that a transferability dispute resolution can be used to detect their dishonesty, for which they can be penalized.

\subsection{Security definitions}

Previously, we introduced the security goals of USS schemes. We are now in a position to define them formally. More than one of the participants can be malevolent, so in general we must look at coalitions of participants that attack the scheme. In an attempt at repudiation, the coalition must include the signer, whereas a coalition aiming to forge a signature does not include the signer. Formally, we define successful cases of repudiation and forging as follows:

{\defn Given a USS protocol $\mathcal{Q}$ and a coalition $C\subset\mathcal{P}$ of malevolent participants -- including the signer $P_0$ -- that output a message-signature pair $(x,\sigma)$, we define \emph{repudiation} to be the function:

\EQ{Rep_C(\mathcal{Q},DR,\sigma,x)=\left\{ \begin{array}{cc}
1 & \textrm{ if } (\sigma,x)\textrm{ is $i$-acceptable for some $i \not\in C$ and $\mathrm{DR}(\sigma,x)=$ \emph{Invalid}} \\
0 & \textrm{ otherwise }
\end{array} \right. }}

Thus, a coalition succeeds at repudiation if they can produce a signature that passes the verification test of one of the honest participants at level $l=0$, but when a $DR$ is invoked, it will be decided that the signature is invalid. According to this definition, a malevolent signer may be able to repudiate with respect to some dispute resolution method, but not other methods.

{\defn Given a USS protocol $\mathcal{Q}$ and a coalition of malevolent parties $C\subset\mathcal{P}$ -- not including the signer $P_0$ -- that output a message-signature pair $(x,\sigma)$, we define \emph{forging} to be the function:

\EQ{Forg_C(\mathcal{Q},\sigma,x)=\left\{ \begin{array}{cc}
1 & \textrm{ if } (\sigma,x)\textrm{ is $i$-acceptable for some $i \not\in C$} \\
0 & \textrm{ otherwise }
\end{array} \right. }}

A successful forgery therefore only requires the coalition to create a signature that passes the verification test of \textit{one} honest participant at level $l=0$. Note that we could have additionally asked that the signature be deemed valid by the $DR$ method, but that would constitute a more difficult task for the attackers.

{\defn Given a USS protocol $\mathcal{Q}$, a coalition of malevolent parties $C\subset\mathcal{P}$ -- including the signer $P_0$ -- that output a message-signature pair $(x,\sigma)$, and a verification level $l$, we define \emph{non-transferability} to be the function:

\EQ{NonTrans_C(\mathcal{Q},\sigma,x,l)=\left\{ \begin{array}{cc}
1 & \textrm{ if } \mathrm{Ver}_{(i,l)}(\sigma,x)=\mathrm{True} \textrm { for some } i\not\in C \textrm{ and } \mathrm{Ver}_{(j,l')}(\sigma,x)=\mathrm{False}\\
 & \textrm{ for some } 0\leq l'<l \textrm{ and some } j\neq i, \textrm{ } j\not\in C\\
0 & \textrm{ otherwise }
\end{array} \right. }}

Therefore, a message-signature pair will be non-transferable at level $l$ if the coalition can produce a signature that at least one honest participant verifies at level $l$, but some other honest participant does not verify at a lower level. Thus, if the signature is non-transferable, there exists a sequence of participants such that, as the signature is transferred in the order of the sequence, at least one of them will not agree that he can transfer the signature to the remaining participants.

We can now state the main security definitions for USS protocols:

{\defn Given a coalition $C\subset \mathcal{P}$, a USS protocol $\mathcal{Q}$ is called \emph{$\epsilon$-secure against forging} if, using their optimal strategy, the probability that the coalition outputs a message-signature pair $(x,\sigma)$ constituting a successful forgery satisfies

\beq
\Pr[Forg_C(\mathcal{Q},\sigma,x)=1]\leq \epsilon,
\eeq

where the probability is taken over any randomness in the generation algorithm and the optimal forging strategy.

}

{\defn Given a coalition $C\subset \mathcal{P}$ and a dispute resolution method $\mathrm{DR}$, a USS protocol $\mathcal{Q}$ is called \emph{$\epsilon$-secure against repudiation} if, using their optimal strategy, the probability that the coalition outputs a message-signature pair $(x,\sigma)$ constituting successful repudiation satisfies

\beq
\Pr[Rep_C(\mathcal{Q},\sigma,x)=1]\leq \epsilon,
\eeq

where the probability is taken over any randomness in the generation algorithm and the optimal repudiation strategy.}

{\defn Given a coalition $C\subset \mathcal{P}$, a USS protocol $\mathcal{Q}$ is called \emph{$\epsilon$-transferable} at level $l$ if, using their optimal strategy, the probability that the coalition outputs a non-transferable message-signature pair $(x,\sigma)$ at level $l$ satisfies

\beq
\Pr[NonTrans_C(\mathcal{Q},\sigma,x,l)=1]\leq \epsilon,
\eeq

where the probability is taken over any randomness in the generation algorithm and the optimal cheating strategy.}

Note that the notion of transferability only makes sense between honest participants. As discussed before, even if the protocol is $\epsilon$-transferable, if a participant transfers a signed message to a dishonest participant, the dishonest person can always deny that they have an assurance of being able to transfer it further. In that case, a transferability dispute resolution method can be invoked at level $l$.

Finally, we note that the security definitions we have provided here can in principle be adapted or relaxed, depending on the particular scope of the protocol. 
For example, depending on the context, it may or may not be useful to be able to cheat with just any recipient, without knowing specifically who this is. For example, a forger may want a message to be accepted specifically by a bank, and it may be of no interest that the message is accepted by another unknown user out of many possible ones. Thus, schemes offering other types of security, such as sufficiently low probability for forging a message with a particular recipient, should not be completely ruled out.

\section{Properties of USS protocols.}\label{Properties}

In this section, we examine several required properties of USS protocols. Understanding these properties is important for several reasons. First, they serve as guiding principles for the construction of new protocols. Additionally, from a fundamental point of view, they provide insight regarding precisely what characteristics of USS protocols give rise to their security. Finally, delineating these properties allows us to construct a coherent picture of the practical challenges to building these protocols as well as their advantages and limitations compared to 
signature schemes with computational security. In the remainder of this section, we list several of these properties and, whenever relevant, prove that they are required for the security of USS protocols.

\begin{obs}
In any secure USS protocol, all classical communication must be authenticated.
\end{obs}

First, authentication is necessary as a guarantee that the participants of the protocol are who they are supposed to be. Otherwise, it would be possible for unauthorized outsiders to participate and compromise the security of the protocol, for example during dispute resolution. Moreover, just as with quantum key distribution, without authentication any USS protocol is subject to a man-in-the-middle-attack, where an attacker impersonates one or more participants, thus rendering the entire scheme insecure. Information-theoretic authentication requires shared secret keys, so the above observation implies that any secure USS protocol requires secret keys shared between the participants, of length proportional to the logarithm of the length of the messages sent \cite{carter79a}.

\begin{obs}\label{ver2}
$\mathrm{Ver}_{(i,l)}(x,\sigma)=\mathrm{True}\Rightarrow Ver_{(i,l')}(x,\sigma)=\mathrm{True} \textrm{ for all } l'< l.$
\end{obs}
Since the verification level of a signature corresponds to the maximum number of times a signature can be transferred, a signature that is verified at a given level should also be verified at all lower levels.

We have mentioned before that in an information-theoretic scenario, it is necessary that each participant has a different verification function with high enough probability. We now show this explicitly, following  
Ref. \cite{swanson2011unconditionally}.

\begin{obs}\label{differentvers}
\cite{swanson2011unconditionally} For any USS protocol that is $\epsilon$-secure against forging, it most hold that 
\beq
\Pr\left(\mathrm{Ver}_{(i,l)}\neq \mathrm{Ver}_{(j,l)}\right)\leq \epsilon
\eeq
for all $l$ and for all $i\neq j$.
\end{obs}
\textit{Proof.} If $\mathrm{Ver}_{(i,l)}= \mathrm{Ver}_{(j,l)}$, then participant $P_i$ can conduct an exhaustive search for a message-signature pair such that $\mathrm{Ver}_{(i,l)}(x,\sigma)=\mathrm{True}$. But since $\mathrm{Ver}_{(i,l)}= \mathrm{Ver}_{(j,l)}$, participant $P_i$ will also have produced a message-signature pair that passes the verification function of participant $P_j$. From observation \ref{ver2}, if participant $P_i$ can produce such a signature, he can also produce a signature such that $\mathrm{Ver}_{(j,0)}(x,\sigma)=\mathrm{True}$, which constitutes successful forging. Therefore, the verification functions must be different at all levels to guarantee security against forging. If the protocol is $\epsilon$-secure against forging, then this should only happen with probability smaller than $\epsilon$.  \qed

Here we should remark that it is possible for probabilistic protocols, in particular quantum signature protocols, to have two participants with the same verification functions, but the probability of this happening must be made small enough for the protocol to be secure. Alternatively, one could consider other security models in which this condition is relaxed. For example, that two participants may have the same verification function with higher probability, but it is unlikely for a cheating party to know who might have the same function. More generally, as further discussed below, there will be conditions not only on the probability that two participants have the same verification function, but also that it should be hard for a participant to guess the verification function of another participant.

\begin{corollary}
A secure USS protocol with a finite number of possible signatures can only exist for a finite number of participants.
\end{corollary}

\textit{Proof.} For a given verification level $l$ and message $x$, a verification function for participant $P_i$ is equivalent to the specification of a subset $S\subset \Sigma$ of signatures such that $\mathrm{Ver}_{(i,l)}(x,\sigma)=\mathrm{True}$. Since the possible number of signatures is a finite set, so is the number of verification functions. From Observation \ref{differentvers}, in any secure protocol, every participant must have a different verification function with high probability, and since there is only a finite number of these functions, there can only be a finite number of participants. \qed 

In principle, one could add new participants to the protocol by using further communication between the new participant and the original ones. Essentially, in order to construct a protocol with $N+1$ participants from a protocol with $N$ participants, the new participant could interact with all others in exactly the same way as if he had participated directly in a protocol with $N+1$ participants. This interaction could happen at a later time than the original distribution stage.

\begin{obs}
The generation algorithm of a secure USS protocol must randomly generate the verification and signing functions.
\end{obs}

\textit{Proof.} If all functions are generated deterministically, then the specification of the protocol is sufficient for every participant to know the signing function and all the verification functions. However, if a participant knows the signing algorithm, forging is trivial since he can produce authentic signatures. Similarly, if a participant knows the verification function of another person, he can conduct an exhaustive search to find a message-signature pair that is validated by the other participant, which constitutes a successful forgery. Finally, if a signer knows the verification function of the other participants, she can conduct an exhaustive search to find a signature that is accepted by one of them at level $l$, but rejected by everyone else at level $l-1$, which allows her to repudiate or break transferability. Thus, a secure protocol requires a randomized generation algorithm.\qed 

The randomness in the protocol may be produced locally by each participant, 
or it may be generated and distributed by a trusted third party. The randomness may arise from the intrinsic randomness of performing measurements on quantum systems, or by other means.
Overall, from the point of view of each participant, the generation algorithm must induce a probability distribution over the possible signing functions as well as the possible verification functions. Therefore, the security of a USS protocol depends crucially on the difficulty of guessing the functions of other participants. We can formalize this requirement with the following observations.

\begin{obs}\label{obsforge}
For a given message $x$, let $S_C$ be the set of signatures that pass the verification functions at level $l=0$ of all members of a coalition $C$. Similarly, let $S_i$ be the set of signatures that pass the verification function at level $l=0$ of a participant $P_i$ outside of the coalition. Then, for any USS protocol that is $\epsilon$-secure against forging, it must hold that
\beq
\frac{|S_i\cap S_C|}{|S_C|}\leq \epsilon \textrm{ for all } i\not\in C,
\eeq
where $|S|$ is the size of a set $S$ and $S_i\cap S_C$ is intersection between $S_i$ and $S_C$.
\end{obs}
\textit{Proof.} Let $(x,\sigma_{c})$ be a message-signature pair drawn uniformly at random from $S_C$. If this signature passes the verification function at level $l=0$ of a participant outside of the coalition, it will constitute a successful forgery. The probability that this happens is given by $\frac{|S_i\cap S_C|}{|S_C|}$, which must be smaller than $\epsilon$ in order for the protocol to be $\epsilon$-secure against forging.
\qed

\begin{figure}
\includegraphics[width=0.7\columnwidth]{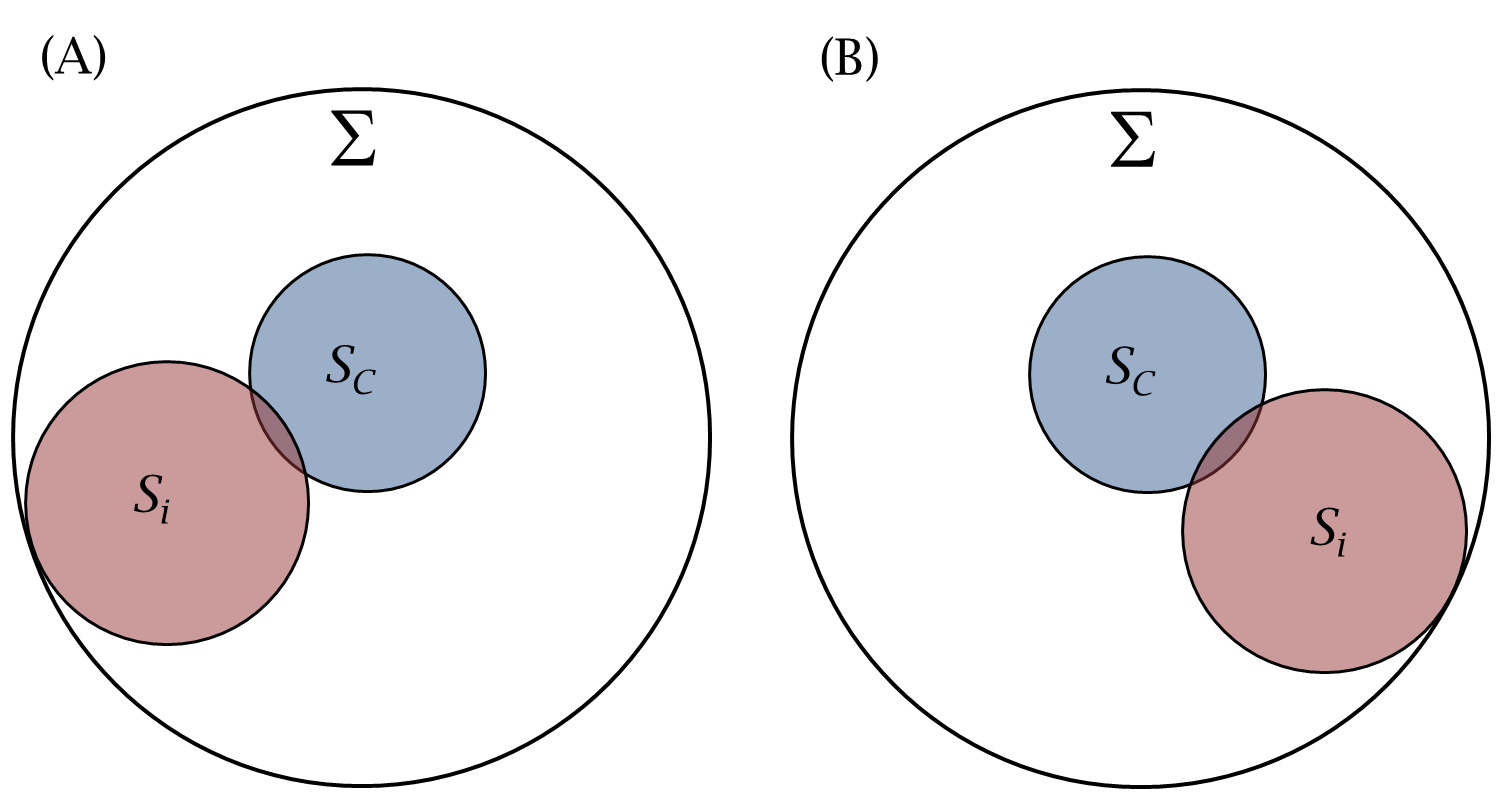}
\caption{$S_C$ is the set of signatures that pass the verification functions at level $l=0$ of all members of a coalition $C$. $S_i$ is the set of signatures that pass the verification function at level $l=0$ of a participant $P_i$ outside of a coalition. If the protocol is secure against repudiation, the intersection $S_C\cap S_i$ must be small compared to $S_C$. More generally, the coalition should not be able to guess 
what the verification functions of the other participants are, except with low enough probability. For example, 
if the protocol is secure against forging, the coalition should not 
be able to distinguish whether they are in situation (A) or (B).}\label{Ver}
\end{figure}

An illustration of the above property can be seen in Figure \ref{Ver}. Notice that if a protocol is correct, authentic signatures are verified by all participants. Therefore, for correct protocols it holds that $S_C\cap S_i\neq \emptyset$. Similarly to the above, we can provide a condition for security against repudiation.

\begin{obs}\label{obsrep}
For a given message $x$, let $S_i$ be the set of signatures that pass the verification function at level $l=0$ of a participant $P_i$ outside of a coalition $C$, and let $\Sigma$ be the set of all possible signatures for this message. Then, for any USS protocol that is $\epsilon$-secure against forging and $\epsilon'$-secure against repudiation with a majority vote dispute resolution, it must hold that
\beq
\frac{|S_i|}{|\Sigma|}\leq \frac{\epsilon'}{1-\epsilon}.
\eeq
\end{obs}
\textit{Proof.} Let $\sigma_r$ be a signature drawn uniformly at random from the set $\Sigma$ of possible signatures. The probability that the signer can repudiate with this signature is given by
\begin{align}
\Pr(Rep)&=\Pr[\mathrm{Ver}_{(i,0)}(x,\sigma_r)=\mathrm{True}\textrm{ AND }\mathrm{MV}(x,\sigma_r)=\mathrm{Invalid}]\nonumber\\
&=\Pr[\mathrm{MV}(x,\sigma_r)=\mathrm{Invalid}|\mathrm{Ver}_{(i,0)}(x,\sigma_r)=\mathrm{True}]\times\Pr[\mathrm{Ver}_{(i,0)}(x,\sigma_r)=\mathrm{True}].\label{EQ:Non-rep condition1}
\end{align} 
If $\sigma_r$ is drawn uniformly at random from $\Sigma$, conditioning on $\sigma_r$ passing the verification function of participant $P_i$ induces a uniform distribution over the set $S_i$. From observation \ref{obsforge}, if the protocol is $\epsilon$-secure against forging, the probability that a signature drawn uniformly at random from $S_i$ passes the verification function of another honest participant must be smaller than or equal to $\epsilon$. Consequently, the probability that a signature drawn randomly from $S_i$ passes the verification function of the \emph{majority} of participants must also be smaller than $\epsilon$, so we have that
\beq
\Pr[\mathrm{MV}(x,\sigma_r)=\mathrm{Valid}|\mathrm{Ver}_{(i,0)}(x,\sigma_r)=\mathrm{True}]\leq \epsilon\nonumber
\eeq 
and therefore
\begin{align}
\Pr[\mathrm{MV}(x,\sigma_r)=\mathrm{Invalid}|\mathrm{Ver}_{(i,0)}(x,\sigma_r)=\mathrm{True}]&=1-Pr[\mathrm{MV}(x,\sigma_r)=\mathrm{Valid}|\mathrm{Ver}_{(i,0)}(x,\sigma_r)=\mathrm{True}]\nonumber\\
&\geq 1-\epsilon.\label{EQ:Non-rep condition2}
\end{align}
If the protocol is $\epsilon'$-secure against repudiation it must hold that $\Pr(rep)\leq \epsilon'$, which, using Eqs. \eqref{EQ:Non-rep condition1} and \eqref{EQ:Non-rep condition2} gives us
\begin{align*}
\epsilon'\geq\Pr(rep)&\geq (1-\epsilon)\Pr[\mathrm{Ver}_{(i,0)}(x,\sigma_r)=\mathrm{True}]\\
&\geq(1-\epsilon)\frac{|S_i|}{|\Sigma|}\\
\Rightarrow &\frac{|\mathrm{Ver}_{(i,0)}|}{|\Sigma|}\leq \frac{\epsilon'}{1-\epsilon},
\end{align*}
where we have used the fact that $\Pr[\mathrm{Ver}_{(i,0)}(x,\sigma_r)=\mathrm{True}]=\frac{|S_i|}{|\Sigma|}$.
\qed

The size of the sets that pass the verification functions at different levels also plays an important role in permitting transferability. In fact, for a special class of USS protocols, such as the QS of Refs. \cite{dunjko2014QDSQKD,dunjko2014quantum,arrazola2014quantum}, it is possible to provide conditions for these sets in order to achieve transferability and security against repudiation. These protocols, which we call \emph{bit-mismatch} protocols, have the following properties. The set of possible signatures $\Sigma$ is the set of all binary strings of $n$ bits, i.e. $\Sigma=\{0,1\}^K$. For each possible message $x$, recipient $P_i$ is given a random subset of positions $p^x_i$ of size $K$ of the integers $\{1,2,\ldots,n\}$. The recipient also receives verification bits $v^x_{i}$. Upon receiving a signature $\sigma$, a recipient collects the bits of $\sigma$ at the positions corresponding to $p^x_i$ to form a shorter string which we call $\sigma_i$. The verification functions are then given by
\beq
\mathrm{Ver}_{(i,l)}(x,\sigma)=\left\{ \begin{array}{cc}
\mathrm{True} & \textrm{ if } h(\sigma_i,v^x_i)\leq s_l K\\
\mathrm{False} & \textrm{ otherwise }
\end{array} \right. 
\eeq
for some $s_l\in[0,\frac{1}{2})$, which depends on the verification level $l$, and where $h(v^x_i,\sigma_i)$ is the Hamming distance between $v^x_i$ and $\sigma_i$.

\begin{obs}
For any correct bit-mismatch protocol which is transferable and secure against repudiation, with a majority-vote dispute resolution method, it must hold that $s_l>s_{l-1}$ for all $l$.
\end{obs}

\textit{Proof.} 
Consider a cheating strategy by the signer in which she randomly flips each bit of the authentic signature $\mathrm{Sign}(x)$ with probability $p$, leading to an altered signature $\sigma'$. For each participant, the choice of $p$ induces a corresponding probability $q_{i,l}(p)$ that the altered signature will pass their verification function at level $l$. Since the protocol is correct, authentic signatures pass the verification functions of all participants at all levels, which implies that $q_{i,l}(0)=1$ and $q_{i,l}(1)=0$ for all $l$. The induced probability $q_{i,l}(p)$ is a continuous function of $p$\footnote{This probability distribution can be shown to be equal to the sum of two cumulative binomial distributions, which are continuous functions.}, which implies that there must exist a value $p_l^*$ such that, for some non-negligible $\delta>0$, it holds that 
\beq
1/2-\delta<q_{i,l}(p_l^*)<1/2\label{EQ: cheat interval}
\eeq
for all participants $P_i$. 

Now consider the case $l=0$ and assume that $s_0\leq s_{-1}$. By choosing $p_0^*$ for her cheating strategy, the signer can create a signature which a given participant accepts with a non-negligible probability greater than $1/2-\delta$ and smaller than $1/2$, according to Eq. \eqref{EQ: cheat interval}. Moreover, since $s_0\leq s_{-1}$, Eq. \eqref{EQ: cheat interval} implies that the probability that any other participant accepts the signature at level $l=-1$ must be smaller than $1/2$. In that case, with non-negligible probability, the majority of participants will reject the signature during dispute resolution, where they check the signature at level $l=-1$. Therefore, such a protocol cannot be secure against repudiation.

Similarly, for the case $l>0$, a dishonest signer can choose $p_l^*$ for her cheating strategy and have any given participant accept a signature at this level with probability at least $1/2-\delta$. If $s_l\leq s_{l-1}$, when the participant who accepts the signature at level $l$ transfers it to another person, the new participant will reject the signature at level $l-1$ with non-negligible probability greater than $1/2$. Thus, such a protocol cannot offer transferability.
\qed

\begin{figure}
\includegraphics[width=0.7\columnwidth]{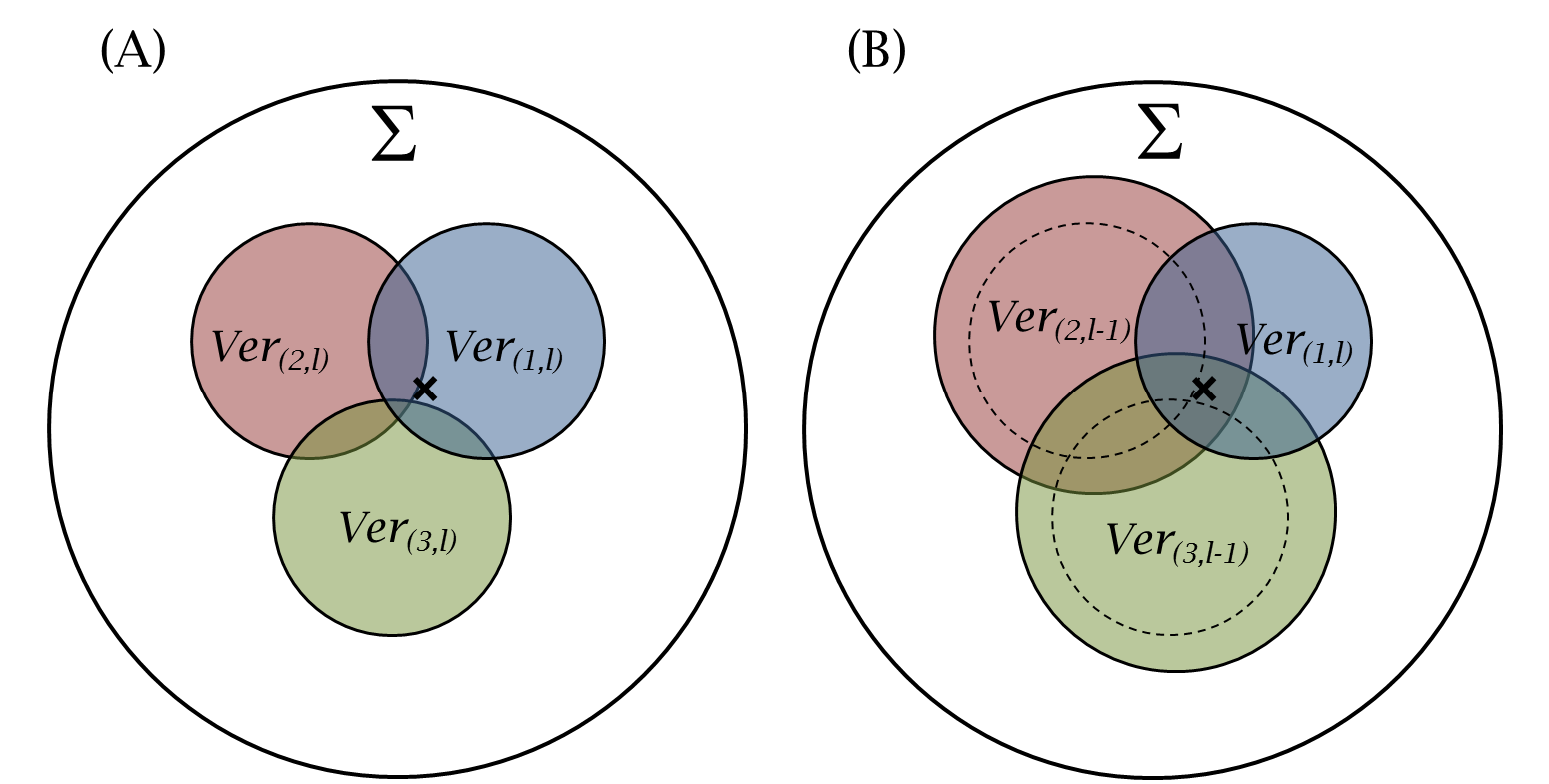}
\caption{For a given verification level $l$, a signer may produce a signature which, with non-negligible probability, passes the verification function at level $l$ of participant $P_1$, but not of the other two participants at this same level. Such a signature is illustrated by a cross in the figure. Since more signatures are accepted at lower levels, when the other participants verify that same signature at level $l-1$, it now passes the verification function of all participants. This feature prevents repudiation and permits transferability. }\label{levels}
\end{figure}

Intuitively, the above proof states that the size of the set of signatures that pass the verification functions at a given level must increase for lower verification levels. This is illustrated in Fig. \ref{levels}.

In the next section, we will examine previous QS protocols in light of our security framework. This will help illustrate our results with concrete examples as well as to showcase the importance of having a rigorous framework.

\section{Previous Protocols}

Here we briefly mention how previous three-party quantum signature protocols fit in our security framework. In particular, we consider the protocol DWA of Ref. \cite{dunjko2014quantum} and the first protocol P1-WDKA from Ref. \cite{dunjko2014QDSQKD}. The experimental realisation in \cite{CollinsQDS} is a variant of the DWA protocol. These two protocols do not require a quantum memory, and thus can be readily compared with classical USS schemes.

In these protocols, we have three participants given by the set $\mathcal{P}=\{P_0,P_1,P_2\}$. The set of possible messages is $X=\{0,1\}$, i.e. we are interested in signing single-bit messages. There is at most one dishonest participant. In the distribution stage, quantum states are exchanged and measured. At the end of this stage, the participants have obtained their verification algorithms. The set of possible signatures for the DWA protocol is $\Sigma=\{0,1\}^K$, while for P1-WDKA it is $\Sigma=\{0,1,2,3\}^K$, where $K$ is the total length of the 
signature.

\subsection{DWA protocol \cite{dunjko2014quantum}}   
\begin{enumerate}
\item All the participants are connected by  
authenticated quantum channels and authenticated classical channels. The assumption of authenticated quantum channels means that the quantum messages which are transmitted are not altered during transmission. This can be guaranteed following a procedure similar to the parameter estimation phase of QKD~
\cite{AWKA15}.
\item For each message $x\in\{0,1\}$, the signer $P_0$ selects a string of bits $\sigma^x$, uniformly at random. For each $0$ in the string $\sigma^x$ he prepares the coherent state $\ket{\alpha}$ and for each $1$ he prepares the coherent state $\ket{-\alpha}$. He then generates this sequence of coherent states twice and sends one copy of this sequence to $P_1$ and the other to $P_2$.
\item The recipients $P_1,P_2$ take their copies and pass them through an optical multiport (see \cite{dunjko2014quantum} for details). The effect of this 
is the following. If all parties are honest, then 
they end up with the state $P_0$ sent, while if there was any deviation on $P_0$'s side  -- for example $P_0$ sending different quantum states to $P_1$ and $P_2$ -- then they end up with a symmetrised quantum state that is identical for both. This step is done to guarantee that the protocol is secure against repudiation. 
\item Finally, each of the recipients measures the received sequence of coherent states $\bigotimes_k\ket{(-1)^{\sigma^x_k}\alpha}$ using unambiguous state discrimination \cite{USD1,USD2,USD3}. The result is that each of the recipients knows the correct bit value for the positions in which he obtains an unambiguous outcome. For participant $i$, we denote the bit string of outcomes as $v^x_{i}$  and the positions for which they obtain unambiguous outcomes as $p^x_{i}$. The recipients have partial knowledge of the signature, but the sender does not know which bits are known to whom, and therefore he will not be able to repudiate.

\item Each participant $P_i$ defines the verification function for the signature $\sigma^x$ as follows. First, they form a shorter string $\sigma_{i}^x$ from $\sigma^x$ by keeping only the bits corresponding to the positions $p^x_{i}$ for which they obtain unambiguous outcomes. The verification function of level $l$ is then defined as 
\beq
\textrm{Ver}_{(i,l)}(x,\sigma)=  \left\{
  \begin{array}{cc}
    \mathrm{True} & \textrm{ if } h(\sigma_{i}^x,v_{i}^x)< s_l K\\
    \mathrm{False} & \textrm{ otherwise }
  \end{array}
\right.
\eeq
where $h(\sigma_{i}^x,v_{i}^x)$ is the Hamming distance between $\sigma_{i}^x$ and $v_{i}^x$, and $s_l$ is a fraction defined by the protocol. Therefore, this protocol is a bit-mismatch protocol, as defined in the previous section. In the original protocol, there were only two thresholds $s_a$ and $s_v$; the first was used to verify whether a signature is transferable and the second to verify just the origin of the signature. In our notation, $s_a=s_1$ and $s_v=s_0$. These fractions satisfy $s_{0}>s_1$.

\item{The signature function is given by $\textrm{Sign}(x)=\sigma_x$.}
\item{Dispute resolution was not explicitly defined. However, it was implicit that a majority vote was to be used.}
\end{enumerate}

Remarks about the security of this protocol will be made after we give the description of the second protocol, since they have several similarities.

\subsection{P1-WDKA protocol \cite{dunjko2014QDSQKD}}

\begin{enumerate}
\item All the participants are connected by  
 authenticated quantum channels and authenticated classical channels.
\item For each message $x\in\{0,1\}$, the signer $P_0$ selects a string $\sigma^x$ of numbers from $\{0,1,2,3\}$, uniformly at random. For $0$ he prepares the qubit state $\ket{0}$, for $1$ the state $\ket{+}=1/\sqrt{2}(\ket{0}+\ket{1})$, for $2$ the state $\ket{1}$ and for $3$ the state $\ket{-}=1/\sqrt{2}(\ket{0}-\ket{1})$. These are usually referred to as the BB84 states. He then generates this sequence of BB84 states twice and sends one copy of this string to $P_1$ and the second copy to $P_2$.

\item For each qubit he receives, recipient $P_1$ ($P_2$) randomly chooses whether to keep this state or forward it to $P_2$ ($P_1$). The effect of this process is that each of the qubits which $P_0$ sends may end up with either of the recipients. In other words, they have now symmetrized the quantum states they have, even if the sender $P_0$ initially deviated and sent different signatures $\sigma^x$ to each one of them. For each message $x$, participant $P_i$ defines the set $p^x_{i}\subset \{1,2,\ldots,K\}$ of positions for which they have a qubit.
\item 
Each of the recipients measures the received BB84 qubits using unambiguous state elimination \cite{USE0,USE}. With this measurement, they \emph{never} learn what state $P_0$ sent -- they only rule out one of the possible states. Therefore, for each position in $p^x_{i}$ for which they had a qubit, they obtain a set of at most two states that are ruled out. The set of states they did not rule out is the set of allowed states, denoted by 
$A^x_{j}\subset \{0,1,2,3\}$ for each position $j$ in $p^x_{i}$. Note that each $A^x_{j}$ has  at least two allowed states but not more than three. We then define the set of $i$-perfect signatures $V^x_i$ as the set that contains all strings of symbols $v^x_i$ where the value of the string for each of the positions in $p^x_i$ is in the set $A^x_{j}$. Again, the recipients have partial knowledge of the signature $\sigma^x$ and the sender $P_0$ is not aware of exactly what this knowledge is or for which positions this information was obtained.

\item Each participant $P_i$ defines the verification function for the signature $\sigma^x$ as follows. First, they form a shorter string $\sigma_{i}^x$ from $\sigma^x$ by keeping only the bits corresponding to the positions $p^x_{i}$ for which they 
received qubits and thus have unambiguously ruled out states. 
The verification function for level $l$ is then defined as 
\beq
\textrm{Ver}_{(i,l)}(x,\sigma)=  \left\{
  \begin{array}{cc}
    \mathrm{True} & \textrm{ if } \min_{v^x_i\in V^x_i} h(\sigma^x_i,v^x_i)
< s_l K\\
    \mathrm{False} & \textrm{ otherwise, }
  \end{array}
\right.
\eeq
where $h(\sigma_{i}^x,v_{i}^x)$ is the Hamming distance between $\sigma_{i}^x$ and $v_{i}^x$, and the minimum is taken over all $i$-perfect signatures. The fraction $s_l$ is again defined by the protocol. In the original protocol, there were two thresholds $s_a$ and $s_v$; the first was used to verify whether a signature is transferable  
and the second to verify just the origin of the signature. In our notation, $s_a=s_1$ and $s_v=s_0$, with $s_{0}>s_1$.

\item{The signature function is given by $\textrm{Sign}(x)=\sigma_x$.}
\item{Dispute resolution was not explicitly defined. However, it was implicit that a majority vote was to be used.}
\end{enumerate}

The full security analysis of these protocols can be found in the original references. However, we here make a few remarks. First, we see that in a certain sense, these protocols are easier to analyse than general multiparty QS protocols because there is at most one dishonest participant. This significantly simplifies proofs for non-repudiation and transferability because the sender $P_0$ cannot have colluding parties. As we will see in the multiparty protocol below, having the sender colluding with recipients can lead to having honest participants totally disagreeing on fractions of the signature, and extra care is needed to address such possibilities. 

Second, these protocols have a property that places strict demands on the noise level and imperfections in an implementation. 
The recipients $P_1$ and $P_2$ receive the same sequence of quantum states. Since they hold a legitimate copy of the state received by the other recipient, they have partial information about the other participant's verification algorithm. This makes it harder to guard against forging, to the point that security is only possible for low levels of noise and experimental imperfections. The security analysis is also complicated by the fact that the optimal forging attack  depends on the states sent, e.g. on the amplitude $\alpha$ of the coherent states.

In any case, the intuition behind the security of these protocols is still the same. Forging is not possible because in order to deceive a participant $P_2$, the other participant $P_1$ should correctly guess 
the bit value for at least a fraction $s_v=s_0$ of the positions in which $P_2$ obtained an unambiguous outcome. Participant $P_1$ can use her copy to make a best guess, but this guess is never perfect, while the unambiguous measurement gives a perfect result when it does give a result, and therefore a legitimate participant always has an advantage. Repudiation in the case of three parties is essentially the same as non-transferability, since the aim is to make one recipient accept at level $l=1$ and the other reject at the lower level $l=0$. The security against this is guaranteed by the fact that the two recipients symmetrize their records, and therefore, from the point of view of $P_0$, he cannot make the one accept a lower threshold and then the other reject a higher threshold.

Finally, it is worth noting that, at the time Refs. \cite{dunjko2014quantum} and \cite{dunjko2014QDSQKD} were written, the security framework we gave here did not exist. Therefore, the concept of dispute resolution and of the extra verification level $l=-1$ 
were not defined. Strictly speaking, for the full security of those protocols, we should define a new level $l=-1$ with threshold $s_{-1}$ that obeys the condition that $s_{-1}>s_0>s_1$. This is another good example of how the framework introduced in this work can prove fruitful in making more accurate statements and even in improving existing protocols.

In the next section, we use the security framework and properties developed so far to generalize the protocol P2-WDKA introduced in Ref. \citep{dunjko2014QDSQKD} to the case of many participants. We provide a full security proof against forging, repudiation and non-transferability.

\section{Generalized P2-WDKA Protocol}\label{Example}

In this protocol, which is a generalization of the protocol P2 of Ref. \cite{dunjko2014QDSQKD}, we have $N+1$ participants given by the set $\mathcal{P}=\{P_0,\cdots,P_{N}\}$. The set of possible messages is $X=\{x_1,\ldots,x_M\}$, where there are $M$ different possible messages. Additionally, $\Sigma=\{0,1\}^K$ is the set of possible signatures, and $K=nN$ is the length of the total signature, where $n$ is an integer that depends on the required security parameters and is divisible by $N$. 

As in any cryptographic protocol, we will make some trust assumptions. In particular, we assume that the number of honest participants\footnote{We assume that the adversaries are static, i.e. the participants are either honest or dishonest for the entire duration of the protocol.} is at least $h$. We can then define the fraction of dishonest participants as $d_f=1-h/N$. The maximum verification level $l_{\max}$ is determined by the allowed fraction of dishonest participants,
\EQ{\label{l_max}(l_{\max}+1)d_f< 1/2.
}
The reason for this restriction will become clear later. The distribution stage of the protocol, which gives rise to the generation algorithm, proceeds as follows:

\begin{enumerate}
\item{All the participants use quantum key distribution links in order to establish pairwise secret keys. Each recipient needs to share a secret key of $nM$ bits with the signer $P_0$ and a secret key of $2\frac{nM}{N}(1+\lceil\log_2 n\rceil)$ bits with each of the other recipients.}
\item{For each possible message $x\in X$, the signer selects a string $\sigma^x$ of $K=nN$ bits uniformly at random and divides it into $N$ sections $\{\sigma^x_1,\sigma^x_2,\ldots, \sigma^x_N\}$. The signer sends $\sigma^x_i$ to participant $P_i$ over a secure channel using their shared secret keys.}
\item{For every possible message, each recipient randomly divides the set $\{1,2,\ldots,n\}$ into $N$ disjoint subsets $\{p^x_{i,1},p^x_{i,2},\ldots,p^x_{i,N}\}$ and uses the bit values of $\sigma^x_i$ at the randomly chosen positions $p^x_{i,j}$ to form the string $v^x_{i,j}$.  }
\item{For all $i\neq j$, each participant $P_i$ transmits the string $v^x_{i,j}$ and the positions $p^x_{i,j}$ to participant $P_{j}$ over a secure channel using their shared secret keys. Participant $P_i$ keeps $v^x_{i,i}$ and $p_{i,i}$ to herself.}
\item{Each participant $P_j$ defines a test for a section $\sigma_i^x$ as follows. First, they form a shorter string $\sigma_{i,j}^x$ from $\sigma^x_i$ by keeping only the bits corresponding to the positions $p^x_{i,j}$. The test is then defined as 
\beq
T^x_{i,j,l}(\sigma_{i}^x)= \left\{
  \begin{array}{cc}
    1 & \textrm{ if } h(\sigma_{i,j}^x,v_{i,j}^x)< s_l\frac{n}{N}\\
    0 & \textrm{ otherwise }
  \end{array}
\right.
\eeq
where $h(\sigma_{i,j}^x,v_{i,j}^x)$ is the Hamming distance between $\sigma_{i,j}^x$ and $v_{i,j}^x$ and $s_l$ is a fraction defined by the protocol. These fractions satisfy
\EQ{\label{s_l}
\frac{1}{2}>s_{-1}>s_{0}>s_1>\cdots > s_{l_{\mathrm{max}}}.
}}
\item{The verification function is defined as 
\beq
\textrm{Ver}_{(i,l)}(x,\sigma)= \left\{
  \begin{array}{cc}
    \mathrm{True} & \textrm{ if } \sum_{j=1}^n T^x_{j,i,l}(\sigma_{j}^x)> N f_l\\
    \mathrm{False} & \textrm{ otherwise }
  \end{array}
\right.
\eeq
where $f_l$ is a threshold fraction given by
\EQ{\label{fraction_threshold}f_l=\frac12+(l+1) d_f.
} 
}
\item{The signature function is given by $\textrm{Sign}(x)=\sigma_x$.}
\item{Majority vote is the dispute resolution method. }
\end{enumerate}

\begin{figure}
\begin{center}
\includegraphics[width=0.6\columnwidth]{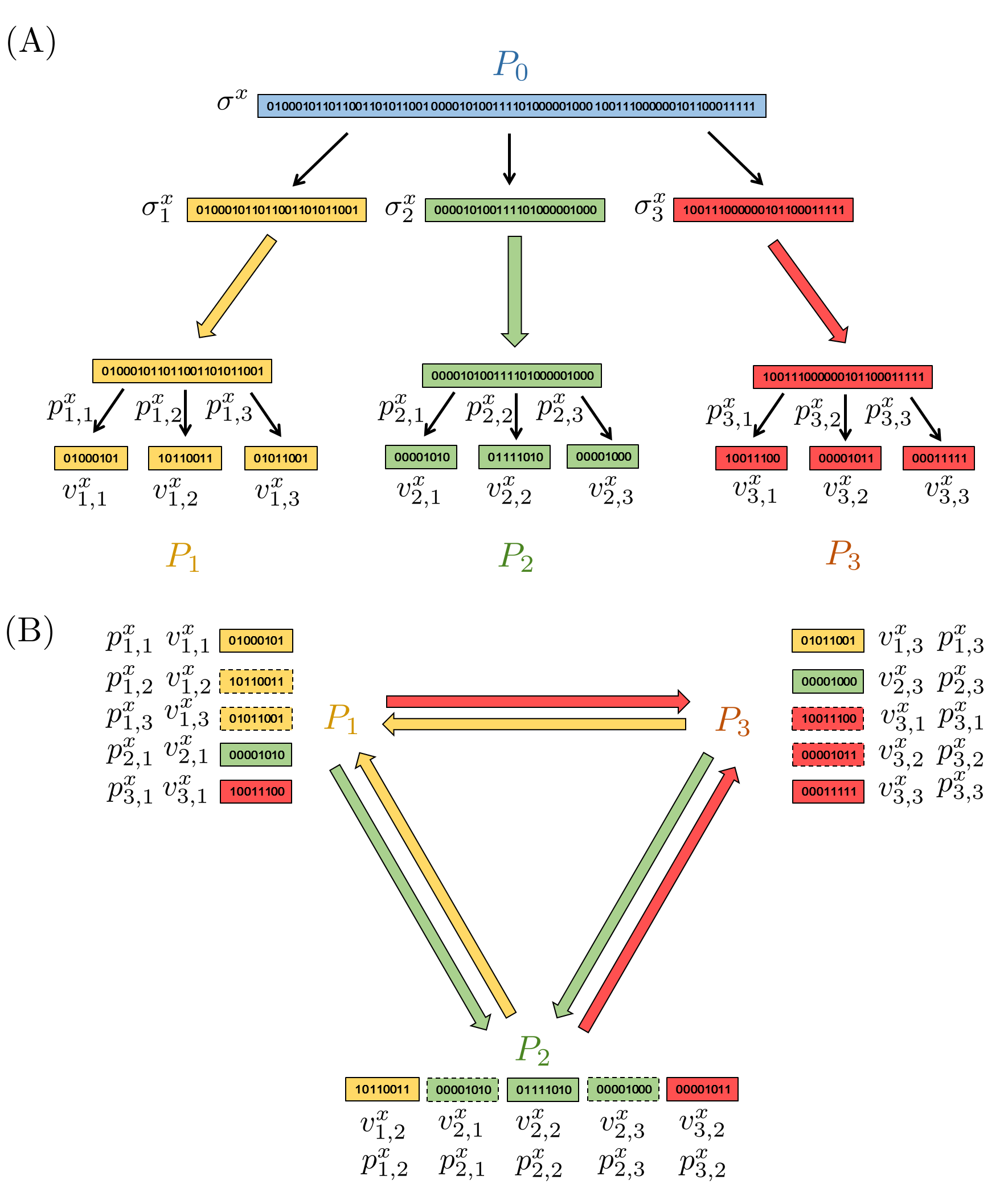}
\caption{Illustration of the protocol with four participants. In part (A), the sender divides a randomly generated string $\sigma^x$ into three sections $\sigma^x_1,\sigma^x_2,\sigma^x_3$ and sends each of them to the corresponding participant over a secure channel, using a secret key previously generated using quantum key distribution. The secret channels are represented by thicker coloured arrows. The other participants divide the sections they receive to produce the strings $v^x_{i,j}$ alongside the corresponding positions $p^x_{i,j}$. In (B), the participants exchange the sections $v^x_{i,j}$ of the signature and the positions $p^x_{i,j}$ over secure channels. In the end, every participant keeps their original sections plus one additional section from each of the other participants, which they use for their verification functions. The sections in dashed boxes are known by the corresponding participant but are not used in the verification functions.}\label{protocol}
\end{center}
\end{figure}

The main steps of the distribution stage are illustrated in Fig. \ref{protocol}.

The verification function, in words, accepts at level $l$ if there are more than a fraction $f_l$ of the sections $\{\sigma^x_1,\sigma^x_2,\ldots, \sigma^x_N\}$ that pass the test of the $i$th participant. This choice of the fraction $f_l$ is made in order to satisfy a few constraints. First, we need the protocol for $l=-1$ to still require more than half of the tests to succeed, i.e. $f_{-1}>1/2$. Second, we want the difference of the thresholds between two levels to exceed the fraction of dishonest participants i.e. $f_l-f_{l-1}>d_f$. Finally, by noting that $f_l\leq 1$ for all $l$, we determine the maximum value that $l$ can take and this results in Eq. \eqref{l_max}.

In the protocol, there are two different types of thresholds, $s_l$ and $f_l$, both depending on the verification level $l$. The first threshold, $s_l$, determines whether a given part of the signature passes the test or not, by checking the number of mismatches at this part. The second threshold, $f_l$, determines how many parts of the signature need to pass the test in order for the signature to be accepted at that level.

An example of why different fractions for each verification level are needed is given by the following. Assume that one recipient, for example $P_1$, is a ``spy'' of an adversarial sender $P_0$, i.e. colludes with her in order to make two honest recipients $P_2$ and $P_3$ disagree on the validity of a signature. The spy can tell the sender the elements $(v^x_{1,2},p_{1,2})$ and $(v^x_{1,3},p_{1,3})$. The sender can then use this information to send a signature $\sigma'$ that differs from the ideal signature $\sigma$ only by flipping all the bit values at the positions determined by $p_{1,3}$. Recipient $P_2$ would accept the message, since he finds no errors. However, $P_3$ will find that \emph{all} the bits of $v_{1,3}$ wrong, which will make his test fail. In general, if $d_f n$ dishonest participants exist, and if all of them are spies, two honest participants can differ by at most $d_f n$ tests. From Eq. \eqref{fraction_threshold}, choosing $f_l-f_{l-1}=d_f$ allows the protocol to remain secure against this type of attack.

Finally, note that the important information defining the verification functions can be encoded in an $n\times n$ matrix, which we call the verification matrix. Each element of this matrix is a collection of $M$ pairs of strings $(v^x_{i,j},p^x_{i,j})$. The strings $v^x_{i,j}$ have length of $\frac nN$ bits, while the position records $p^x_{i,j}$ have length $\frac{n}{N}\times\lceil\log_2 n\rceil$ bits. Note that in $v^x_{i,j}$ and $p^x_{i,j}$, the first index corresponds to the section $\sigma^x_i$ received by participant $P_i$, while the second index determines the other participant $P_j$ with whom this string is shared. Importantly, it is not mandatory that these verification functions are constructed following the same steps as in the distribution stage outlined above. The security of the protocol relies only on the properties of the verification matrix and the value of other protocol parameters, which in principle may be generated by other means e.g. with the help of a trusted arbiter.

If the participants were honest during the above distribution stage, we end up exactly with the outcome of the ideal generation algorithm, which gives rise to the desired verification and signing functions. The important thing to notice is that deviating in the distribution stage is equivalent to being honest at this stage, but deviating at a later stage of the protocol. The sender gains nothing by sending a different signature to the recipients during the distribution stage, since this is equivalent to sending the correct signature during the distribution stage, but then sending a different signature at a later stage. The same holds for an adversarial recipient who is in coalition with the sender. On the other hand, an adversarial recipient $P_i$ who wishes to forge a message by deviating and giving different $(v^x_{i,j},p^x_{i,j})$, is not improving his chances to forge, since in order to forge a signature for participant $P_1$ for example, he will have to guess correctly the $(v^x_{i,1},p^x_{i,1})$ and even if he is honest, 
he knows the $(v^x_{1,2},p^x_{1,2})$.

We now proceed to prove the security of this protocol. In the following, for simplicity, we will drop the superscript labelling the message $x$ from $v^x_{i,j}$, $p^x_{i,j}$ and $T^x_{(i,j,l)}$, and we will refer to participants by their index only, i.e. as $i$ instead of $P_i$.

\subsection{Security proofs} We will separately address the security of this protocol against forging, repudiation and non-transferability.
We begin by noticing that the value of $n$ must be chosen depending on other parameters and on the level of security. In particular, we want the probabilities for forging, non-transferability, and repudiation to decrease exponentially fast with $n$. However, the number of participants $N$ also enters the security expressions. To make sure that the all cheating probabilities go to zero even when the number of participants is very large, in general we require that
\EQ{\label{N_n}
n\geq\alpha N^{1+\delta},
}where $\alpha\gg 1$ is a large positive constant and $\delta$ a small positive constant.

\textbf{Forging.}  In order to forge, a coalition $C$ which does not include the signer needs to output a message-signature pair $(x,\sigma)$ that is $i$-acceptable for some $i\notin C$. In general, according to our definitions, we consider forging successful if the coalition can deceive \emph{any} honest participant, and not a fixed one. Here, for simplicity, we restrict attention to trying to deceive a fixed participant, and we will prove that this probability decays exponentially fast with the parameter $n$. At the end, we will extend this to the general case where the target is not a fixed participant. Therefore, for now, we fix the recipient that the coalition wants to deceive to be simply $i$. 

Recall that a signature $\sigma$ is $i$-acceptable if $\mathrm{Ver}_{(i,0)}(x,\sigma)=\mathrm{True}$. By the definition of the verification functions of our protocol, this means that the coalition should output a signature $\sigma$ such that participant $i$ accepts $N f_0$ tests at level zero, $T_{j,i,0}(\sigma^j)$. From Eq. \eqref{fraction_threshold}, we have that $f_0=\frac12 +d_f$. By the definition of the protocol, the number of members in a coalition is at most $Nd_f$. The coalition knows the pairs $(v_{j,i},p_{j,i})$ for all $j\in C$, so they can use this knowledge to trivially pass $N d_f$ tests. It follows that in order to forge, the coalition must pass at least $N(f_0-d_f)=\frac N 2$ tests out of the $N(1-d_f)$ tests that they do not have access to. The first step to compute the probability that they can do this is to calculate the probability of passing a single test $T_{j,i,0}$ for $j\notin C$.

\begin{enumerate}

\item We denote the probability to pass a test at level $l=0$ for a coalition with no access to the pair $(v_{i,j},p_{i,j})$ by $p_{t}$. Because the strings $(v_{i,j},p_{i,j})$ were transferred over secure channels by honest recipients, they are completely unknown to the coalition and hence the probability of guessing correctly a single bit of $v_{i,j}$ is exactly $1/2$. In order to pass the test, the coalition needs to guess at least a fraction $s_0$ of bits out of a total of $\frac nN$ bits. The probability that they can achieve this can be bounded using Hoeffding's inequality as
\EQ{p_t\leq \exp\left(-2(1/2-s_0)^2\frac nN\right),
} 
which decays exponentially with the number $\frac nN$ provided that $s_0<1/2$. Note that, from by Eq. \eqref{N_n}, we know that this term decays exponentially even for $N\rightarrow\infty$.

\item Now we will give a bound for the probability of forging against a fixed participant. This can be obtained by computing the probability of passing at least one of the unknown $N(1-d_f)$ tests, which is given by 
\begin{align}
\Pr(\mathrm{Fixed Forge})&< 1-(1-p_{t})^{N(1-d_f)}\approx N(1-d_f)p_t\nonumber\\
&\leq (1-d_f)N\exp\left(-2(1/2-s_0)^2\frac nN\right),\label{forge}
\end{align}
where we have used the fact that $p_t\ll 1$ in the approximation. Again, this probability goes to zero exponentially fast in the parameter $n$. Note also that, by Eq. \eqref{N_n}, this expression goes to zero even for the case $N\rightarrow\infty$, as the term with $p_t$ goes exponentially fast to zero while the other term grows only linearly in $N$.

\item We have now computed the probability to deceive a fixed participant $i$. The total number of honest participants is $N(1-d_f)$ and for successful forging we require that any one of them is deceived. We therefore obtain 

\EQ{\Pr(\mathrm{Forge})=1-(1-\Pr(\mathrm{FixedForge}))^{N(1-d_f)}\lesssim N^2(1-d_f)^2\exp\left(-2(1/2-s_0)^2\frac nN\right).}

\end{enumerate}

\noindent\textbf{Transferability.} In order to break the transferability of the protocol, a coalition $C$ which includes the signer $P_0$ must generate a signature that is accepted by recipient $i\notin C$ at level $l$, while rejected by another recipient $j\notin C$ at a level $l'<l$. To provide an upper bound, we allow for the biggest coalition $C$ that includes $Nd_f$ participants, i.e. all the dishonest participants. For simplicity, again we will fix the participants whom the coalition is trying to deceive to be the $i$th and $j$th, while all the other honest participants are labelled with the index $k$. In general, according to our definitions, transferability fails if the coalition forms a signature that is not transferable for \emph{at least one} pair of honest participants $i,j$. Therefore, we should take into account all possible pairs of honest participants. Here, we first focus on the case of a fixed pair of participants, and we give at the end the more general expressions. The members of the coalition $C$ are labelled with the index $c$.

We first give a sketch of the proof. The first step is to compute $p_{m_{l,l'}}$. This is the probability that the tests corresponding to a part of the signature $\sigma^k$ of an honest recipient $k$ satisfy the following conditions: (i) The test $T_{k,i,l}$ of an honest recipient $i$ at level $l$ is passed \emph{and} (ii) the test $T_{k,j,l'}$ of another honest recipient $j$ at a level $l'<l$ is failed. The second step of the proof is to prove that in order for non-transferability to be successful, there must be at least one test corresponding to an honest participant which the two recipients $i$ and $j$ disagree on. The third step is to combine the previous two steps to provide a bound for the probability of non-transferability for a fixed pair of recipients. Finally, we can use the previous results to bound the probability of non-transferability for any pair of honest recipients.

\begin{enumerate}

\item First, we compute $p_{m_{l,l'}}$, which is the probability that the $k$th test $T^{x}_{k,i,l}$ of an honest recipient $i$ at level $l$ is accepted \emph{and} the test  $T^{x}_{k,j,l'}$ of another honest recipient $j$ at a level $l'<l$ is rejected. The relevant part of the signature is $\sigma^k$, where $k$ is an honest recipient. The two parts of the verification matrix that are relevant are $(v_{k,i},p_{k,i})$ and $(v_{k,j},p_{k,j})$. Since the sender is in the coalition, they know the values of all the sections $v_{i,j}$, but they are completely ignorant of the positions $p_{k,i}$ and $p_{k,j}$, since participants $k,i$ and $j$ are all honest. The coalition can decide to send signatures in such a way that they introduce an average fraction of mistakes $p_e$ compared to the ideal signature that was used to generate the verification algorithms. Thus, the average fraction of mistakes is under their control. Since the protocol is symmetric for all participants, this average fraction of mistakes will be the same for all honest participants and in particular for both $i$ and $j$.

To compute a bound on the joint probability of $i$ accepting at level $l$ and $j$ rejecting at level $l-1$ we will consider
\begin{align}
p_{m_{l,l'}}&=\Pr\left(i\textrm{ accepts at level $l$ AND }j\textrm{ rejects at level $l'$}\right)\nonumber\\
&\leq \min\{\Pr\left(i\textrm{ accepts at level $l$}\right),\Pr\left(j\textrm{ rejects at level $l'$}\right)\}.
\end{align}
The probability of passing the test at level $l$ with an average error $p_e$ can be bounded using Hoeffding's inequalities to be below $\exp \left(-2(p_e-s_l)^2\frac nN\right)$. This is the case since the expected number of mistakes are $\frac nN p_e$ while the mistakes that are tolerated for acceptance are $\frac{n}{N}s_l$. We note that this expression holds for $p_e>s_l$. However, as we will see, for $p_e<s_l$ the probability for the participant rejecting at level $l'<l$ will be even smaller, and since for our bound we consider the minimum of those two probabilities, we can assume that $p_e>s_l$.

The probability of failing the test at level $l'$ with average errors $p_e$ can similarly be bounded to be smaller than $\exp \left(-2(s_{l'}-p_e)^2\frac nN\right)$. This is since the expected mistakes are $\frac nN p_e$ while the mistakes needed to fail are more than $s_{l'} \frac nN$. We note, that this expression holds for $p_e<s_{l'}$. However, as we have seen, for $p_e>s_{l'}$, the probability for the participant accepting at level $l$ will be even smaller (recall $s_l<s_{l'}$), and since for our bound we consider the minimum of those two probabilities, we can assume that $p_e<s_{l'}$.

Therefore, the coalition must choose a value of $p_e$ satisfying
\beq
s_{l}<p_e<s_{l-1}.
\eeq
Since we are taking the minimum over both cases, the best choice for the coalition is to have both probabilities coincide. This is achieved by using a fraction of errors $p_e=(s_l+s_{l-1})/2$ and in that case we obtain the bound
\EQ{
p_{m_{l,l'}}\leq \exp\left(-\frac{(s_{l'}-s_{l})^2}2\frac {n}{N}\right)
}
which decays exponentially with $\frac nN$ and it also depends on the difference $(s_{l'}-s_l)$.

\item It is trivial for the coalition to make two recipients disagree in any way they wish for the results of a test that involves a member of the coalition, i.e. they can make $T^{x}_{c,i,l}$ and $T^{x}_{c,j,l'}$ take any values they wish. However, the number of those tests are at most $Nd_f$, which is the maximum number of members in the coalition.

For the participant $i$ to accept a message at level $l$, he needs a fraction greater than $f_l$ of the tests to pass at this level. On the other hand, for the participant $j$ to reject the message at level $l'$, a fraction greater than $1-f_l'$ of tests must fail at this level. Therefore, even taking the best case for the coalition, which is $l'=l-1$, since it holds that $f_l=f_{l-1}+d_f$, in order for the non-transferability to be successful, the honest participants $i$ and $j$ need to disagree on at least $Nd_f+1$ tests. As we saw, the coalition can easily make them disagree on the $N d_f$ tests originating from them, but the participants $i$ and $j$ still have to disagree on at least one more test originating from an honest participant.

\item In order for the coalition to successfully cheat, the number of tests that pass for the $i$th recipient must be at least $Nf_l+1$. Out of those tests we can assume that $Nd_f$ were due to the coalition, but there are still $N(f_l-d_f)+1$ tests that the coalition does not have access to. In order for the non-transferability to be successful, at least one of these $N(f_l-d_f)+1$ tests should fail for participant $j$ at level $l'=l-1$. The probability that they agree in all of them is $(1-p_{m_{l,l'}})^{N(f_l-d_f)+1}$ and therefore the probability for fixed non-transferability can be bounded as
\begin{align}
\Pr(\mathrm{FixedNonTrans})&\leq 1-(1-p_{m_{l,l'}})^{N(f_l-d_f)+1}\nonumber\\
&\approx \left[N(f_l-d_f)+1\right] p_{m_{l,l'}}\nonumber\\
&\leq \left[N(f_l-d_f)+1\right]\exp\left(-\frac{(s_{l'}-s_{l})^2}2\frac{n}{N}\right)+O(p_{m_{l,l'}}^2).
\label{NonTrans}
\end{align}
This goes to zero exponentially with $\frac nN$. Note that the first term scales linearly in $N$, but $p_{m_{l,l'}}$ decays exponentially with $\frac nN$, therefore with the choice of Eq. (\ref{N_n}) this probability also vanishes at all limits of interest.

\item Finally, we should consider the general case, where the participants $i,j$ are not fixed. Again, we can see that because the probability for fixed parties decays exponentially in the parameter $n$, the protocol remains secure. The number of honest pairs of participants is $[N(1-d_f)][N(1-d_f)-1]/2:=N_p$, so we obtain
\EQ{\Pr(\mathrm{NonTrans})=1-(1-\Pr(\mathrm{FixedNonTrans}))^{N_p}\approx O(N^3)\exp\left(-\frac{(s_{l'}-s_{l})^2}2\frac{n}{N}\right).
}

\end{enumerate}

\noindent\textbf{Non-repudiation.} In order to repudiate, a coalition $C$ including the sender $P_0$ generates an $i$-acceptable signature for some $i\notin C$, where invoking the dispute resolution $DR$ results in $\mathrm{Invalid}$. This means that the coalition wants to make any participant accept a signature at level $l=0$, but then have the majority of participants  to reject the same signature at level $l=-1$. We can actually reduce this problem to the special case of non-transferability from level $l=0$ to level $l=-1$ in the following three steps.

\begin{enumerate}

\item We first find the probability of non-transferability for a fixed pair of participants, i.e. from a fixed honest participant $i$ at level $l=0$ to another fixed honest participant $j$ at level $l=-1$. We denote this probability by $p_1$ and, as found before, it can be bounded by
\EQ{
p_1\lesssim \left[N(f_0-d_f)+1\right] p_{m_{0,-1}}\leq\left(\frac{N}{2}+1\right)\exp\left(-\frac{(s_{-1}-s_0)^2}2\frac n{N}\right),
}
where we have used the fact that $(f_0-d_f)=\frac 12$ from Eq. \eqref{fraction_threshold}.

\item The second step is to note the following. For a fixed recipient $i$ to accept at $l=0$, it means that at least $Nf_0+1=N(\frac 12 +d_f)+1$ parts of his signature were accepted. Out of these, $\frac N2+1$ must have come from honest participants. Now, each of those honest participants that sent $i$ a part that passed his tests also sent the other honest participants sections which, with probability $1-p_1$, pass their tests at level $l=-1$. For a message to be declared invalid in the dispute resolution $DR$, half of the participants have to reject. However, at least $\frac N2+1$ are unlikely to reject, since the probability that they do reject is $p_1$, which can be made arbitrarily small. In other words, for the $DR$ to give $\mathrm{Invalid}$, at least one of the honest participants needs to fail the transferability for a fixed pair of participants. 

\item It is now clear that if no fixed pair of honest participants $i,j$ fails the transferability for levels $l=0$ to $l=-1$, then the coalition cannot repudiate. This leads to the following bound for the probability of repudiation,
\begin{align}\label{repudiation}
\Pr(\mathrm{Rep})&\leq 1-(1-p_1)^{N_p}\approx N_p p_1+O(p_1^2)\nonumber\\
&\leq  O(N^3)\exp\left(-\frac{(s_{-1}-s_0)^2}2\frac n{N}\right),
\end{align}
where $N_p$ as before is the number of honest pairs $[N(1-d_f)][N(1-d_f)-1]/2$ and $p_1$ decays exponentially with $\frac nN$. 

\end{enumerate}

We have seen that all security parameters, from Eqs. \eqref{forge}, \eqref{NonTrans} and \eqref{repudiation}, go to zero exponentially fast with $\frac nN$, provided correct choices of $s_l$ and $f_l$ are made. As stressed before, by Eq. \eqref{N_n}, we also know that these parameters go to zero even if the number of participants $N$ goes to infinity.

\noindent\textbf{Secure channels from QKD.} Security proofs for quantum key distribution (QKD) rely on the assumption that the parties wishing to exchange a secret key behave honestly. In the context of our multiparty protocol for quantum signature schemes, this assumption does not hold, since some of the participants performing QKD may be dishonest. However, we can show that this does not present a problem for the security of our protocol in three steps. Similar arguments are made in \cite{AWKA15}.

\emph{Step 1: Only honest-dishonest QKD links may be affected.} The first observation is that dishonest behaviour during QKD may only be an issue when the QKD link connects an honest participant with a dishonest one. 
For two honest participants, standard QKD security proofs apply, so we are not concerned with this scenario. For the case of two dishonest participants, since all members of the coalition have access to the same information -- as is assumed in our security definitions -- it is irrelevant whether they behave honestly during QKD. Similarly, honest participants do not eavesdrop on dishonest participants, so there are no consequences to the security of the QS protocol.

In the following two steps we will show that for the case of  an honest and a dishonest participant using QKD to establish a shared secret key, \emph{any} adversarial behaviour during the QKD stage of the protocol is equivalent to a dishonest behaviour in subsequent parts of the protocol. Therefore, we can assume that the participants were honest during the QKD stage and examine all possible deviations for later stages of the protocol.

\emph{Step 2: No-gain from leaking information.} At the end of a QKD protocol, an honest participant $P_i$ holds a key register $X$ which, in the ideal case, is identical to the string $Y$ of the other participant $P_c$ and is completely unknown to any other party. This means that any dishonest behaviour by participant $P_c$ can only lead to two possible outcomes: (i) The registers $X$ and $Y$ are not identical, or (ii) $X$ is correlated with the register of another party. Since we assume that all dishonest participants are in coalition, all of them have perfect knowledge of the register $Y$, so there is no need to eavesdrop information about this string. They of course benefit from knowledge of $X$, but they can have perfect knowledge of $X$ simply if $P_c$ behaves honestly during QKD. Therefore, leakage of information does not help the adversarial coalition.

\emph{Step 3: No-gain from imperfect keys.} Similarly, if there are mismatches between the registers $X$ and $Y$, any message which is transmitted secretly by using a one-time pad with either $X$ or $Y$ will be received with errors in all positions in which $X$ and $Y$ differ. However, if $Y$ is used by $P_c$ to transmit a message to the honest participant $P_i$, the situation is exactly equivalent to one in which they have identical secret keys, but $P_c$ decided to introduce errors in the message sent to $P_i$. Similarly, if $P_i$ is the one sending the message, the situation is equivalent to the keys being identical but participant $P_c$ introducing errors after receiving the message. In fact, since in order to cheat, the coalition needs to know the verification function of the honest participants, their optimal strategy is to be honest during the QKD stage and have a perfect copy of the other participants' secret keys. Therefore, the security of QKD is only relevant in a quantum signature scheme in order to protect honest participants who want to establish a secret key. It is precisely in this regime that standard QKD proofs apply.

\section{Discussion}\label{Discussion}

In this work, we have provided a full security framework for quantum signature schemes. We have generalized the security definitions of Swanson and Stinson \cite{swanson2011unconditionally} to allow for quantum schemes and different levels of verification. Additionally, we have proven several properties that USS protocols, quantum or classical, must satisfy in order to achieve their security goals. Together, these results form a powerful set of tools to be employed in the understanding and development of improved protocols in a general setting.

In fact, we have done just that by using our security framework to generalize the P2-WDKA protocol of Wallden et. al \cite{dunjko2014QDSQKD} to the multiparty case. This protocol is secure against forging, repudiation, and non-transferability, relying on minimal security assumptions. Interestingly, the quantum-mechanical features responsible for the security of the protocol can be completely outsourced to quantum key distribution (QKD), where a vast literature of sophisticated security proofs already exists. This feature also addresses the issue of authentication in quantum signature schemes: we can simply use QKD to generate new secret keys to be used in the authentication of future instances of a signature protocol. Finally, since this protocol can be implemented using any point-to-point QKD network, it is already practical, making experimental demonstrations in the short-term future a real and exciting possibility.

As a consequence of our results and those of Ref. \cite{dunjko2014QDSQKD}, the status of unconditionally secure signature schemes should be considered analogous to that of secure communication, where a classical protocol -- the one time-pad -- already exists and can guarantee information-theoretic security at the expense of shared secret keys. Quantum communication can then be used to establish these secret keys via unsecured quantum channels. Similarly, for signature schemes, there exist classical protocols -- such as our generalized P2-WDKA protocol -- which provide information-theoretic security at the expense of shared secret keys. Remarkably, even in the setting where parties are dishonest, quantum key distribution can be used to establish the secret keys. Overall, we can now understand unconditionally secure signature schemes as a practical application of quantum key distribution. Future work can focus on optimizing these classical protocols, for example in reducing the length of the secret keys that need to be exchanged as a function of the message size. Additionally, it is important to continue  
to study protocols where  
quantum communication can be used to construct quantum signature schemes without the need to distil a secret key. Those schemes could offer advantages in terms of scalability, or in terms of extending the distance between parties, and thus be proven more useful in this respect.  
For example, Ref. \cite{AWKA15} discusses how such ``direct quantum" signature schemes may be practical even if the quantum bit error rate is too high to allow the distillation of a secure key.

\acknowledgments

The authors would like to thank A. Ignjatovic, N. L\"{u}tkenhaus and  D. Stinson for valuable discussions. In particular, the authors would like to thank V. Dunjko for discussions in our previous collaboration, especially \cite{dunjko2014QDSQKD}, where the three-party protocol P2-WDKA that was generalised in the present paper was invented, mainly by him. This work was supported by Industry Canada, the NSERC Strategic Project Grant (SPG) FREQUENCY, the NSCERC Discovery Program and the UK Engineering and Physical Sciences Research Council (EPSRC) under EP/K022717/1 and EP/M013472/1. J.M. Arrazola is grateful for the support of the Mike and Ophelia Lazaridis Fellowship. P.W. gratefully acknowledges support from the COST Action MP1006.

\bibliography{Refs}
\bibliographystyle{unsrt}
 \end{document}